# Large spin Hall effect in 5d-transition metal anti-perovskites

Priyamvada Jadaun*, Leonard F. Register, and Sanjay K. Banerjee
Department of Electrical and Computer Engineering, The University of Texas at Austin, Austin, TX 78712

**Abstract**
The spin Hall effect (SHE) is highly promising for spintronic applications, and the design of materials with large SHE can enable ultra-low power memory technology. Recently, 5d-transition metal oxides have been shown to demonstrate a large SHE. Here we report large values of SHE in four 5d-transition metal anti-perovskites which makes these anti-perovskites promising spintronic materials. We demonstrate that these effects originate in the mixing of $d_{x2-y2}$ and $d_{xy}$ orbitals caused by spin orbit coupling.

**Introduction**

Spintronic devices that exploit both the electrical charge and spin of an electron are highly promising for next-generation memory technology [1]. Along with their non-volatility and energy efficiency for memory applications, spintronic devices are notable for their compatibility with conventional CMOS circuits, high endurance, and fast random access [2, 3, 4]. A prominent spintronic memory device is the spin Hall effect (SHE)-based magnetic random-access memory (MRAM) [5]. The SHE is a phenomenon by which an electrical current, when passed through certain nonmagnetic materials produces a pure spin polarized current [6, 7]. SHE-based MRAMs utilize this spin current to switch the magnetic state of memory bits, leading to all-electrical control of memories. However, energy efficient implementation of SHE based MRAMs requires the discovery of spin Hall materials with a giant spin Hall efficiency [1].

Recently, 5d-transition metal oxides have attracted significant attention as promising spin Hall materials [8]. The large spin orbit coupling (SOC) of the 5d-transition metal atom along with the tunability of transition metal oxides by Fermi level makes these materials particularly attractive. Experimental results have demonstrated large spin Hall values in $IrO_2$ [8] and $SrIrO_3$ [9], and theoretical calculations have predicted 5d-transition metal anti-perovskites as being particularly promising for large spin Hall values owing to their weak crystal field [10]. Here, we report large SHE for four new 5d-transition metal anti-perovskites $Ir_3WC$, $Pt_3LaP$, $Ta_3PbO$ and $Ir_3PbO$ that are comparable to the large SHE values predicted for $SrIrO_3$. We also note that the sign of SHE is opposite for $Ta_3PbO$ and $Ir_3PbO$, as expected based on Hund's rules, given the orbital filling [11]. We report that a significant contribution to SHE arises from the mixing of the $d_{x2-y2}$ and $d_{xy}$ orbitals caused by spin orbit coupling. These results demonstrate that 5d-transition metal anti-perovskites are a promising class of spin Hall materials. Further studies aimed at enhancement of SHE via Fermi level tuning could achieve even larger SHE values [10].

*priyamvada@utexas.edu

**Structure and calculation details**

An anti-perovskite $A_3BC$, has the structure of an inverted perovskite in which the perovskite cations and anion exchange sites. In such a structure, a transition metal atom, when placed on site A, experiences weak crystal field as it is bonded only to two ligand atoms (C). This weak crystal field makes anti-perovskites a promising class of large SHE materials [10]. Here, we calculate SHE in four anti-perovskites. Structures for these materials were taken from reference [12] for $Ir_3WC$ and [13] for $Pt_3LaP$. To obtain the structures for $Ta_3PbO$ and $Ir_3PbO$, we started with an initial structure for $Yb_3PbO$ from reference [14] and substituted Yb with Ta and Ir, respectively. All structures were then optimized to find the minimum energy lattice parameters. Calculations of spin Hall conductivity were perform using QUANTUM ESPRESSO [15], WANNIER90 [16], and our in-house code. The in-house code interfaces with the former two codes and calculates the SHE from their output. In the past, we have successfully used this method for the prediction of SHE values [11]. We carried out density functional theory (DFT) using QUANTUM ESPRESSO to obtain the electronic ground states for the materials utilizing a plane-wave energy cut off of 50 Ry and a k-mesh of 15×15×15.

We utilized norm-conserving, fully relativistic LDA pseudopotentials [17, 18] that were constructed using the atomic pseudopotential engine [19]. We benchmarked these pseudopotentials against the fully relativistic all-electron potential. Subsequently, we used WANNIER90 to map our DFT ground-state wave functions onto a maximally localized Wannier function basis and employed an adaptive k-mesh strategy to extract the matrices relevant to the calculation of SHC.

**Results**

Results of spin Hall conductivity are tabulated in Table 1. In general, the materials show a large SHE that is comparable to the SHE values predicted for $SrIrO_3$ [10]. These results are consistent with the rational design principles derived previously by the authors and point to 5d-transition metal oxides as being promising for spintronic applications. Further enhancement of SHE in these materials would require engineering of the Fermi level placement, say via doping or design of oxide heterostructures.

| Material | Spin Hall conductivity $\sigma_{\alpha\beta}^s$ ($\hbar/2e$ S/cm) |
|---|---|
| $Ir_3WC$ | 503.6 |
| $Pt_3LaP$ | 535.0 |
| $Ta_3PbO$ | -407.2 |
| $Ir_3PbO$ | 281.0 |

Table 1. Spin Hall conductivity in 5d-transition metal anti-perovskites.

## Ir₃WC

Fig. 1 shows plots of SHE in Ir$_3$WC. We notice peaks of spin Berry curvature (shown in Fig. 1(b)) around X, Y and Z that are expected to contribute significantly to the spin Hall conductivity. The projected band structure (shown in Fig. 1(c) & (d)) demonstrates considerable orbital mixing between $d_{x2-y2}$ and $d_{xy}$ orbitals, which we expect contributes to the large SHE.

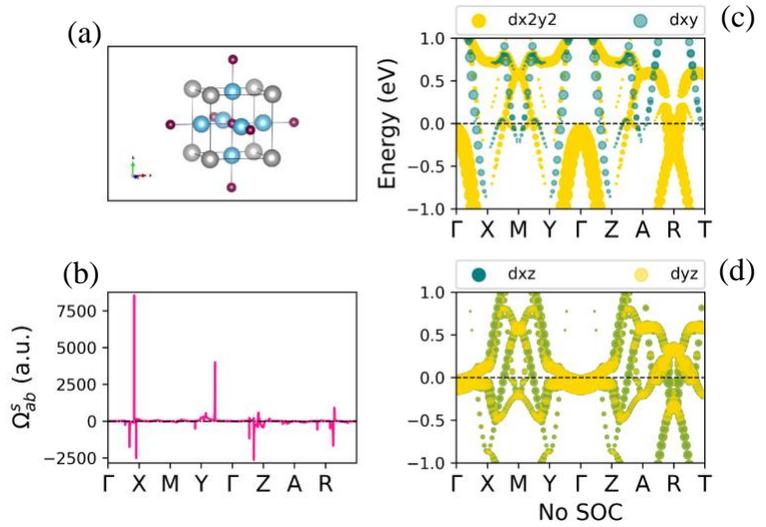

Figure 1. SHE in Ir$_3$WC. (a) shows the structure with Ir in blue, W in grey and C in maroon, (b) demonstrates the spin Berry curvature, and (c) and (d) plot the band structure projected onto d orbitals, with $d_{x2-y2}$ and $d_{yz}$ in yellow and $d_{xy}$ and $d_{xz}$ in green, respectively.

## Pt₃LaP

Fig. 2 demonstrates the SHE plots for Pt$_3$LaP with the spin Berry curvature plot shown in Fig. 2 (b). Spin Berry curvature demonstrates a noticeable peak at the M point and smaller peaks around the A point. Comparing this with the projected band structure plotted in Fig. 2. (c), these points are clearly regions of band crossing between $d_{x2-y2}$ and $d_{xy}$ bands, which signifies the importance of this contribution towards the large SHE seen in Pt$_3$LaP.

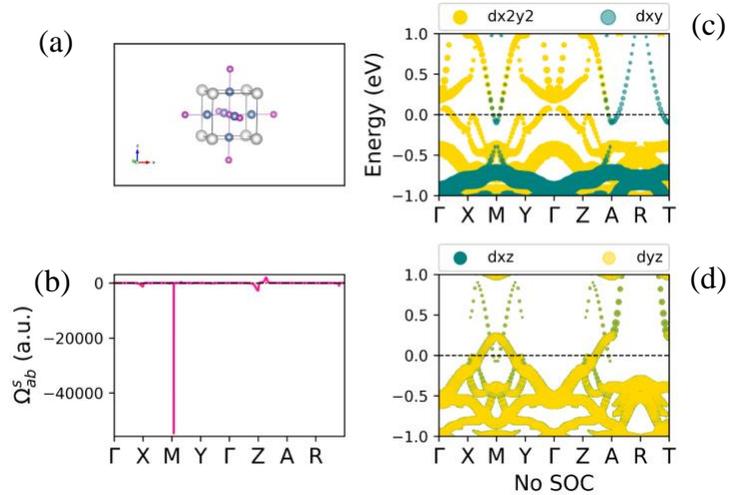

Figure 2. SHE in Pt$_3$LaP. (a) shows the structure with Pt in blue, La in grey and P in pink, (b) demonstrates the spin Berry curvature, and (c) and (d) plot the band structure projected onto d orbitals, with $d_{x2-y2}$ and $d_{yz}$ in yellow and $d_{xy}$ and $d_{xz}$ in green, respectively.

## Ta$_3$PbO

Plots for SHE in Ta$_3$PbO are shown in Fig. 3. The large peak in spin Berry curvature seen in Fig. 3 (b) corresponds to the d$_{x2-y2}$ band, suggesting a likely contribution from this band towards SHE.

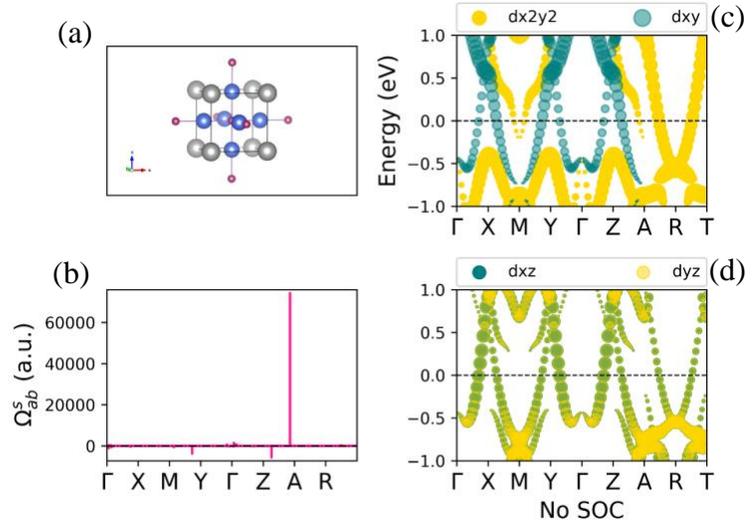

Figure 3. SHE in Ta$_3$PbO. (a) shows the structure with Ta in blue, Pb in grey, and O in pink, (b) demonstrates the spin Berry curvature, and (c) and (d) shows the band structure projected onto d orbitals, with d$_{x2-y2}$ and d$_{yz}$ in yellow and d$_{xy}$ and d$_{xz}$ in green, respectively.

## Ir$_3$PbO

Fig. 4 demonstrates SHE plots for Ir$_3$PbO. We note a large spin Berry curvature around the A point (see Fig 4 (b)), which coincides very well with the overlap between the d$_{x2-y2}$ and d$_{xy}$ bands. (See Fig. 4 (c).) We expect the mixing of d$_{x2-y2}$ and d$_{xy}$ orbitals to be a dominant contribution to SHE in Ir$_3$PbO.

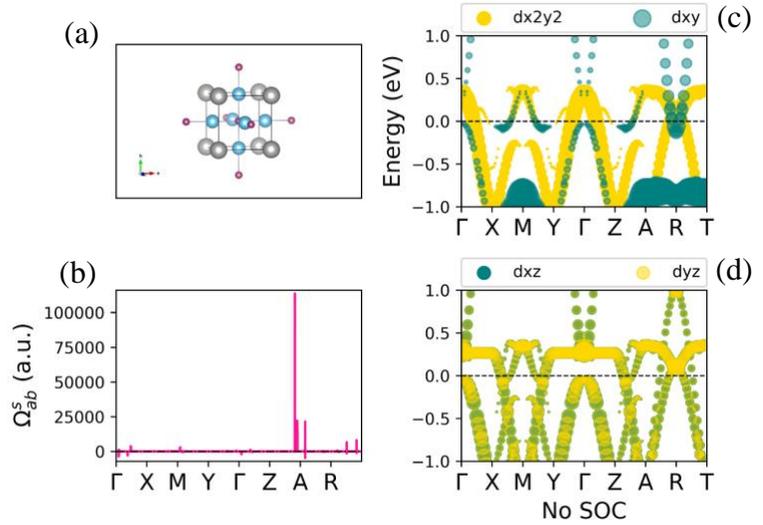

Figure 4. SHE in Ir$_3$PbO. (a) shows the structure with Ir in blue, Pb in grey and O in pink, (b) demonstrates the spin Berry curvature, and (c) and (d) shows the band structure projected onto d orbitals, with d$_{x2-y2}$ and d$_{yz}$ in yellow and d$_{xy}$ and d$_{xz}$ in green respectively.

## Conclusion

We have presented a computational study of SHE in 5d-transition metal anti-perovskites. These materials have been previously predicted to be promising for exhibiting large SHE. We report large calculated values of SHE for $Ir_3WC$, $Pt_3LaP$, $Ta_3PbO$ and $Ir_3PbO$ that are comparable to the large SHE values predicted for $SrIrO_3$ previously. We also report that a major contribution to SHE in these materials originates from the overlap between the $d_{x2-y2}$ and $d_{xy}$ orbitals. Further enhancement of SHE values could be achieved by fine tuning the placement of the Fermi level in these materials. 5d-transition metal oxides are, thus, a promising class of materials for obtaining large spin Hall effects.

## Acknowledgments

We thank NSF for financial support under grant NNCI-2025227 and NSF EFRI-newLAW grant 1802167. We also acknowledge the Texas Advanced Computing Center at The University of Texas at Austin (https://www.tacc.utexas.edu) for providing HPC resources that have contributed to the research results reported within this paper.

## Data Availability Statement

The data that support the findings of this study are available from the corresponding author upon reasonable request.

## References

[1] W. Yan, E. Sagasta, M. Ribeiro, Y. Niimi, L. E. Hueso and F. Casanova, *Nat. Comm.,* vol. 8, no. 661, 2017.

[2] C. Chappert, A. Fert and F. N. Van Dau, *Nat. Mater.,* no. 6, p. 813–823, 2007.

[3] S. A. Wolf, D. D. Awschalom, R. A. Buhrman, J. M. Daughton, S. von Molna, M. L. Roukes, A. Y. Chtchelkanova and D. M. Treger, *Science,* vol. 294, p. 1488–1495, 2001.

[4] G. A. Prinz, *Science,* vol. 282, p. 1660–1663, 1998.

[5] L. Liu, C.-F. Pai, D. C. Ralph and R. Buhrman, *Phys. Rev. Lett.,* vol. 109, no. 18, p. 186602, 2012.

[6] M. I. Dyakonov and V. I. Perel, *Phys. Lett. A,* vol. 35, p. 459–460, 1971.

[7] J. E. Hirsch, *Phys. Rev. Lett.,* vol. 83, p. 1834–1837, 1999.

[8] K. Fujiwara, Y. Fukuma, J. Matsuno, H. Idzuchi, Y. Niimi, Y. Otani and H. Takagi, *Nat. Comm.,* vol. 4, p. 2893, 2013.

[9] T. N. e. al., *PNAS,* vol. 116, no. 33, p. 16186–16191, 2019.

[10] P. Jadaun, L. F. Register and S. K. Banerjee, *PNAS,* vol. 117, no. 22, pp. 11878-11886, 2020.

[11] N. Reynolds, P. Jadaun, J. T. Heron, C. L. Jermain, J. Gibbons, R. Collette, R. A. Buhrman, D. G. Schlom and D. C. Ralph, *Phys. Rev. B,* vol. 95, p. 064412, 2017.

[12] D. V. Suetin, I. R. Shein and A. L. Ivanovskii, *Solid State Sciences,* vol. 12, p. 814–817, 2010.

[13] H. Chen, X. Xu, C. Cao and J. Dai, *Phys. Rev. B,* vol. 86, p. 125116, 2012.

[14] K. Persson, "Materials Data on Yb3PbO (SG:221) by Materials Project. DOI: 10.17188/1188128," 2014.

[15] P. G. e. al., *J. Phys.: 769 Condens. Matter,* vol. 21, p. 395502, 2009.

[16] A. A. Mostofi, J. R. Yates, Y.-S. Lee, I. Souza, D. Vanderbilt and N. Marzari, *Comput. Phys. Commun.,* vol. 178, no. 685, 2008.

[17] J. P. Perdew and Y. Wang, *Phys. Rev. B,* vol. 45, p. 13244, 1992.

[18] J. P. Perdew and A. Zunger, *Phys. Rev. B,* vol. 23, p. 5048, 1981.

[19] M. J. T. Oliveira and F. Nogueira, *Comp. Phys. Comm.,* vol. 178, pp. 524-534, 2008.

*priyamvada@utexas.edu